\newcommand{\bm}[1]{\mbox{\boldmath$#1$}}

\documentclass[twocolumn,showpacs,preprintnumbers,amsmath,amssymb]{revtex4}

\usepackage{graphicx}
\usepackage{dcolumn}
\usepackage{bm}

\begin{document}

\title{ Garbay-Toulouse Phase Transition in Heisenberg Spin-Glasses \\
in Three Dimensions} 

\author{T. Shirakura, D. Ninomiya$^1$, Y. Iyama$^1$, and F. Matsubara$^1$} 

\affiliation{Faculty of Humanities and Social Sciences, Iwate University,
Morioka 020-8550, Japan\\ 
$^1$Department of Applied Physics, Tohoku University, Sendai 980-8579, Japan}

\date{ \today }

\begin{abstract}

We examine three-dimensional $\pm J$ Heisenberg models with and without 
random anisotropies in a magnetic field. 
We calculate both the stiffness exponent $\theta_s$ at absolute zero 
temperature and spin-glass correlation lengths for the longitudinal 
and transverse spin components at finite temperatures. 
We suggest that, contrary to a chirality scenario predicted by Kawamura 
and his co-workers, a Gabay-Toulouse phase transition occurs when the 
anisotropies are absent, although no phase transition occurs 
when they are present. 

\end{abstract}


\maketitle

Phase transitions of three-dimensional (3D) Heisenberg spin-glass (SG) models 
have attracted much interest in recent years. 
Two phase-transition scenarios are involved in controversy. 
One is the {\it SG scenario}, in which a usual SG phase transition takes place 
at a finite temperature. However, this scenario has been believed to hold only 
when anisotropies are present\cite{Iyota}. 
The other is {\it a chirality-glass (CG) scenario}, proposed by 
Kawamura and his co-workers\cite{Kawamura1,Kawamura2,Hukushima1}. 
In the CG scenario, not the spins, but the local 
chiralities freeze at a finite temperature. 
In this scenario, the SG phase transition never occurs 
in isotropic SG models. 
The freezing of the spins was suggested to occur through coupling of 
the spins and the local chiralities by random anisotropies. 


Kawamura and his co-workers gave three pieces of evidence of the CG scenario 
in the isotropic case: 
the stiffness exponent for the chiralities is positive $\theta_{\chi} > 0$, 
whereas that for the spins is negative $\theta_{s} < 0$\cite{Kawamura1}; 
only the chirality autocorrelation exhibits 
a pronounced aging effect at low temperatures\cite{Kawamura2}; 
the chirality overlap distribution $P(q_{\chi})$ exhibits a one-step-like 
replica symmetry breaking (RSB) behavior\cite{Hukushima1}. 
However, reexaminations of those properties revealed different aspects: 
the stiffness exponent for the spins in a lattice with open boundaries 
is positive $\theta_s > 0$ \cite{Matsubara1,Endo1,Matsubara2}; 
the spin autocorrelation of a system, in which a uniform rotation is 
removed, exhibits an aging effect similar to that of the chirality 
autocorrelation\cite{Rotat,BerthierYoung}; 
and the SG susceptibility $\chi_{\rm SG}$ exhibits a divergence behavior 
to a finite temperature $T_{\rm SG}$\cite{Matsubara2}. 
Recently, the SG phase transition temperature $T_{\rm SG}$ and the CG 
phase transition temperature $T_{\rm CG}$ were estimated using a 
non-equilibrium relaxation method\cite{Nakamura1} and the scaling method 
of the SG correlation length\cite{LeeYoung}. Results suggested 
that $T_{\rm SG}\;=\;T_{\rm CG}$ in both methods. 
Based on those results, the SG scenario has come back also in the 
isotropic models. 
That is, a usual SG phase transition takes place at a finite temperature. 
Freezing of the local chiralities results. 


The controversy surrounding those two scenarios has reached a new stage. 
It has been speculated that true SG properties are visible only in large 
lattices, {\it e.g.}, the $L \times L \times L$ lattice with $L\gtrsim20$, 
because the coupling of the spins and the local chiralities, 
which exists even for the isotropic model in small lattices, 
loosens for $L \rightarrow \infty$\cite{Hukushima2}. 
Campos {\it et al.} quite recently studied the model 
for big lattices ($L =$ 24 and 32)\cite{Campos} to resolve this issue. 
Their results suggest that the lower critical dimension $d_l$ of this model 
is equal to or a slightly smaller than 3 ($d_l \lesssim 3$) and 
that a large finite size correction exists in the scaling property 
of the model with $d = 3$. 
Having taken into account this correction, they also suggested that 
$T_{\rm SG} = T_{\rm CG}$. 
However, objections exist in relation to their interpretation\cite{Campbell}. 
Unfortunately, it is too difficult to resolve this issue herein. 


Two scenarios predict different aspects for a finite magnetic field 
$H \neq 0$. 
In the SG scenario, a usual phase transition will take place, 
which is characterized by a freezing of the transverse component of the spin, 
i.e., a Garbay-Toulouse (GT) phase transition\cite{G_T}. 
In the CG scenario, the CG phase transition will occur, but 
the SG phase transition is absent\cite{Kawamura3}. 
More interesting is a case in which anisotropies are present. 
In the SG scenario, the SG phase transition will disappear 
because of a random field effect\cite{Imry_Ma}. 
On the other hand, Imagawa and Kawamura predicted that the CG transition 
still occurs at $H \neq 0$, accompanied with the SG phase 
transition\cite{Imagawa}.

In this letter, we present an examination of the phase transition of 
the $\pm J$ Heisenberg models with and without random anisotropies 
at a finite magnetic field $H \neq 0$. 
Special attention is devoted to an induced magnetic moment 
$\langle \bm{S}_i\rangle$ at each site $i$. We consider a SG spin component, 
$\tilde{\bm{S}}_i (\equiv \bm{S}_i - \langle \bm{S}_i\rangle)$, to examine 
cooperative phenomena of the system. 
Results show that, in the isotropic model, the ground state stiffness and 
the scaling property of the SG correlation length suggest 
the presence of the SG (GT) phase transition, like those at $H = 0$. 
On the other hand, in the anisotropic model, both the CG transition 
and the SG transition disappear. 
Therefore, we suggest that a usual SG phase transition occurs in the isotropic 
Heisenberg SG model at $H \neq 0$ as well as at $H = 0$.


We study the $\pm J$ Heisenberg SG models in three dimensions ($d = 3$) 
in a magnetic field $H$ described using the Hamiltonian:
\begin{eqnarray}
   \mathcal{H} = - \sum_{\langle\it{ij}\rangle}\it{J}_{\it{ij}}
	\bm{S}_{\it{i}}\bm{S}_{\it{j}}  
     -\sum_{\langle\it{ij}\rangle}\sum_{\mu\nu}D_{ij}^{\mu\nu}S_i^\mu S_j^\nu
     - \it{H} \sum_i S_i^z,
\end{eqnarray}
where $\bm{S}_i$ is the classical vector spin of $|\bm{S}_i|=1$; 
$J_{ij}=+J$ or $-J$ with the same probability of 1/2. 
The second term expresses the anisotropic energy; 
$D_{\it{ij}}^{\mu\nu} (= D_{\it{ji}}^{\mu\nu} = D_{\it{ij}}^{\nu\mu})$ 
($\mu,\nu=x,y,z$) are symmetric random anisotropic constants 
distributed in the range $[-D:D]$. 
The lattice is a simple cubic lattice of $L \times L \times (L+1)(\equiv N)$ 
with periodic or skew boundary conditions along two $L$ directions 
and a periodic boundary condition along the $(L+1)$ direction. 
We consider two cases: (A) $D = 0$ and (B) $D \neq 0$.


\vspace{0.5cm}
\noindent
(A) Isotropic case of $D = 0$


First, we consider the ground state stiffness of the model 
using a method proposed by Matsubara {\it et al.}\cite{Matsubara1}. 
Here we consider lattices of $L \times L \times (L+1)$ with open boundaries 
for the $(L+1)$ direction. 
The lattice has two opposite surfaces $\Omega_1$ and $\Omega_{L+1}$. 
We first determine the ground state spin configuration $\{\bm{S}_i\equiv 
\bm{S}_i^{\parallel}+\bm{S}_i^{\perp}\}$ and its energy $E_L^0$. 
Then, fixing all the spins on the surface $\Omega_1$, 
all the spins on the surface $\Omega_{L+1}$ are rotated by 
the same angle $\phi = \pi/2$ around the $z$-axis and fixed.  
Under this boundary condition, we calculate the minimum energy of the system, 
$E_L^{\phi}$, which is always higher than $E_L^0$. 
The stiffness of the system might be characterized by the excess energy 
$\Delta E_L (\equiv E_L^{\phi}-E_L^0)$. The stiffness exponent 
$\theta_s$ might be defined by the relation $\Delta E_L \propto L^{\theta_s}$.

We have calculated [$\Delta E_L$] of the model up to $L = 12$, 
together with the parallel (the magnetization) and the transverse 
components of the spins, $[M^z(=|\sum_i\bm{S}_i^{\parallel}|/N)]$ and 
$[S^{\perp}(=\sum_i|\bm{S}_i^{\perp}|/N)]$, having used a genetic 
algorithm\cite{Baba}, in which $[\cdots]$ means a sample average. 
Numbers of samples with different bond distributions are $100 \sim 1000$. 
Figure 1 shows $[M^z]$ and $[S^{\perp}]$ as functions of $H$. 
Those values depend little on $L$, 
suggesting that they are those for $L \rightarrow \infty$. 
In fact, $[M^z]$ exhibits a characteristic property of the SG, 
i.e., it increases rapidly with $H$ and saturates gradually 
at high magnetic fields $H_s \sim 7J$. 
Consequently, $[S^{\perp}]$ has a considerable value up to $H_s$. 
Figure 2 shows [$\Delta E_L$] for several $H$ in a Log-Log form. 
Using least-squares fitting, we estimated the stiffness exponent 
as $\theta_s = 0.58\pm0.02, 0.65\pm0.02, 0.63\pm0.03, 0.63\pm0.06, and 
\  -0.18\pm0.15$, respectively, for $H/J = 0, 2, 4, 6, and \  7$. 
It is interesting that $\theta_s$ are positive and almost equal for $H < H_s$. 
This result is analogous to that in the spin-flop (SP) phase of 
an antiferromagnetic Heisenberg (AFH) model, in which $\theta_s = 1$ 
for $H < H_s (\equiv H_{\rm C}=12 J)$. 
Therefore, we expect that the SG (GT) phase transition occurs at 
$H < H_{\rm C}(= 6.5\pm0.5 J)$.

\begin{figure}[tbhp]
\begin{center}
\includegraphics[width=5.5cm,clip]{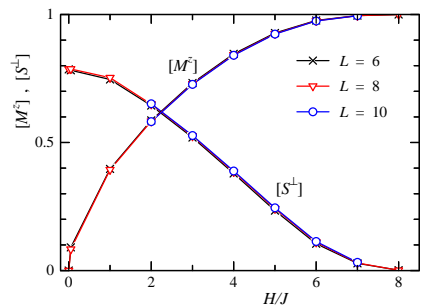}\\
\end{center}
\vspace{-0.4cm}
\caption{
Magnetization $[M^z]$ and the transverse component $[S^{\perp}]$ 
for various lattice with $L$ as functions of $H$. 
}
\label{fig:1}
\end{figure}

\begin{figure}[tbhp]
\begin{center}
\includegraphics[width=6.5cm,clip]{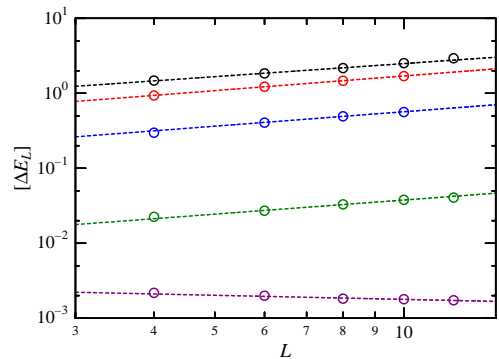}\\
\end{center}
\vspace{-0.4cm}
\caption{(Color online) 
Excess energy $[\Delta E_L]$ for various lattices with $L$. The symbols 
denote, from the above, those at $H/J = 0, 2, 4, 6 \  and \  7$. 
}
\label{fig:2}
\end{figure}

We next examine the phase transition of the model. 
We consider a two-replica system with 
$\{\bm{S}_i^{\alpha}\}$ and $\{\bm{S}_i^{\beta}\}$. 
At $H \neq 0$, the spins are polarized to the $z$-direction: 
$m_i (\equiv \langle S_i^z \rangle^{\alpha} = 
\langle S_i^z \rangle^{\beta}) \neq 0$, where $\langle \cdots \rangle$ 
means a thermal average. 
Magnitudes of $m_i$ will vary from site to site. 
Figure 3 shows their distribution $P(m_i)$. 
In fact, $m_i$ distribute in a very wide range. 
The slight size dependence of $P(m_i)$ reveals 
that $m_i$ is a purely magnetic-field induced one. Then we subtract 
$\bm{m}_i(\equiv (0,0,m_i))$ from the original spin\cite{LeeYoung}: 
%
$  \tilde{\bm{S}}_i^{\alpha,\beta} = \bm{S}_i^{\alpha,\beta} - \bm{m}_i. $
%
Hereafter, we call $\tilde{\bm{S}}_i^{\alpha,\beta}$ SG components 
and consider their cooperative phenomena.

We consider the spin-glass correlation lengths $\xi_L^{\eta}$ for the 
longitudinal ($\eta=\;\parallel$) and transverse ($\eta=\perp$) components. 
We calculate them using a formula\cite{LeeYoung}: 
\begin{equation}
  \xi _L^{\eta} =\frac{1}{2\sin{(\bm{k}_{\mathrm{min}}/2)}}
   \left(\frac{\tilde{\chi}^{\eta}_{SG}(\bm{0})}{\tilde{\chi}^{\eta}_{SG}
   (\bm{k}_{\mathrm{min}})} -1\right)^{1/2},
   \label{teigi.xi}
 \end{equation} 
where $\bm{k}_{\mathrm{min}}=\left(0,0,2\pi /(L+1)\right)$. 
The $\bm k$-dependent SG susceptibility is given as 
 $ \tilde{\chi}^{\parallel}_{SG}(\bm{k})=N\left[\langle |\tilde{q}^{zz}(\bm{k})
 |\rangle^{2}\right]$,  
and  $ \tilde{\chi}^{\perp}_{SG}(\bm{k})=N\sum_{\mu,\nu=x,y}
\left[\langle |\tilde{q}^{\mu\nu}(\bm{k})| \rangle^{2}\right]$, 
with
$\tilde{q}^{\mu\nu}(\bm{k})=\frac{1}{N}\sum_{i}\tilde{S}_{i}^{\alpha\mu}
\tilde{S}_{i}^{\beta\nu}\exp{(i\bm{k}\cdot\bm{R_{i}})}$. 
If a SG phase transition occurs, the correlation length divided by the 
system size $L$, $\xi_L^{\eta}/L$, has the following scaling property:
\begin{equation}
  \frac{\xi_L^{\eta}}{L}
  =\hat{\xi}^{\eta}\left(L^{1/\nu}(T-T_{\rm SG}(H))\right), 
\end{equation}  
where $\nu$ is the correlation length exponent, $T_{\rm SG}(H)$ is 
the transition temperature at $H$, 
and $\hat{\xi}^{\eta}$ represents a scaling function.

\begin{figure}[ttt]
\begin{center}
\includegraphics[width=5.5cm,clip]{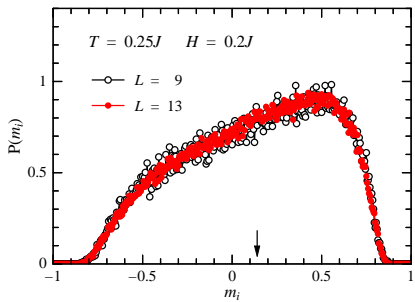}\\
\end{center}
\vspace{-0.4cm}
\caption{(Color online) 
Distribution of the site magnetizations $m_i$ of the model with $D = 0$ 
in a finite magnetic field. $T = 0.25 J$ is slightly higher 
than $T_{\rm SG} (\sim 0.2 J)$ at $H = 0$. 
The arrow indicates the average 
value of $m_i$ for $L = 13$. 
}
\label{fig:3}
\end{figure}

We performed a simulation of this two replica system on the lattice 
with $L \leq 23$ having used a temperature exchange MC 
method\cite{Hukushima3} with an over-relaxation\cite{Campos}. 
Numbers of samples are 128 for the largest lattice; equilibration 
is checked by monitoring the stability of the results against runs 
at least twice as long. 
Figures 4(a) and 4(b) respectively show $\xi_L^{\perp}/L$ 
and $\xi_L^{\parallel}/L$ at $H = 0.2J$ as functions of $T$. 
These two quantities exhibit different size dependence. 
$\xi_L^{\perp}/L$ for different $L$ cross around $T = 0.22 J$, suggesting 
the presence of the phase transition. 
In fact, choosing $T_{\rm SG} \sim 0.215J$, we can scale 
$\xi_L^{\perp}$ (see Fig. 5) using a finite size correction exponent of 
$\phi = 0.9$ predicted by Campos {\it et al.}\cite{Campos}. 
This result is compatible with the ground state study. 
In contrast, $\xi_L^{\parallel}/L$ seem not to cross, 
even at low temperatures, suggesting the absence of the phase transition 
in the longitudinal component. 
This result is also compatible with that of the Ising SG model at $H \neq 0$, 
where no-AT line was suggested\cite{Young_Ising_h}.

We have considered properties of the isotropic SG model at $H \neq 0$. 
Both the ground state stiffness and the scaling properties of the SG 
correlation length suggest the occurrence of the SG phase transition 
in the transverse component of the spins. 
That is, the GT phase transition will occur at $H \neq 0$.

\begin{figure}[tbhp]
\begin{center}
\includegraphics[width=6.5cm,clip]{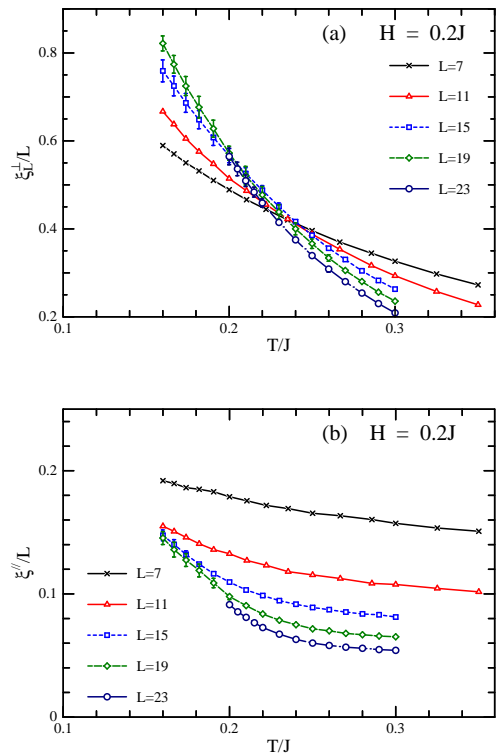}\\
\end{center}
\vspace{-0.4cm}
\caption{ (Color online) 
Plots of (a) $\xi_L^\perp /L$ and (b) $\xi_L^\parallel /L$ of the model 
with $D=0$.
}
\label{fig:4}
\end{figure}

\begin{figure}[tbhp]
\begin{center}
\includegraphics[width=5.5cm,clip]{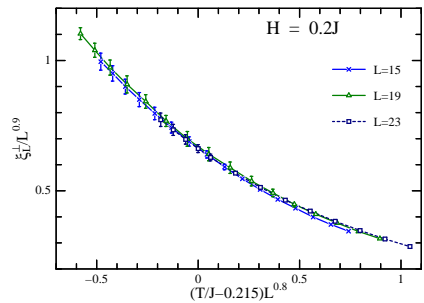}\\
\end{center}
\vspace{-0.4cm}
\caption{ (Color online) 
A finite-size scaling plot of $\xi_L^{\perp}$. 
}
\label{fig:5}
\end{figure}


\vspace{0.3cm}
\noindent
(B) Anisotropic case of $D \neq 0$

Next we consider the anisotropic model with $D = 0.1J$. 
We first note that the system has only the spin reversal symmetry of 
$\{\bm{S}_i\} \rightarrow \{-\bm{S}_i\}$ at $H = 0$.
This symmetry breaks at $H \neq 0$. 
Consequently, {\it the system at $H \neq 0$ will exhibit no phase transition 
associated with the spins}. 
On the other hand, Imagawa and Kawamura (IK)\cite{Imagawa} examined the CG 
phase transition and suggested that the CG phase transition occurs because of 
the one-step-like RSB accompanied with the SG phase transition. 
They considered cooperative phenomena of the original spins 
$\{\bm{S}_i^{\alpha}\}$ and $\{\bm{S}_i^{\beta}\}$. 
We have reexamined it\cite{Comm_MC} using SG components 
$\{\tilde{\bm{S}}_i^{\alpha}\}$ and $\{\tilde{\bm{S}}_i^{\beta}\}$. 
Figure 6 shows $\xi_L^\perp /L$ as functions of $T$. In stark contrast 
to the case of $D = 0$, the $\xi_L^\perp /L$ for different $L$ seem not 
intersect at any finite temperature. 
This result supports the former argument.

Does the CG phase transition really occur at $H \neq 0$? 
We have also reexamined the chirality transition using the SG components 
$\{\tilde{\bm{S}}_i^{\alpha}\}$ and $\{\tilde{\bm{S}}_i^{\beta}\}$. 
Figure 7 shows the chirality overlap distribution of 
$P(q_{\chi})$\cite{Comm_q} at a low temperature. 
In marked contrast to the IK results (see Fig. 8 in 
\cite{Imagawa}), $P(q_{\chi})$ exhibits a single peak at $q_{\chi} = 0$, 
which becomes sharper as $L$ increases. 
This result indicates no freezing of the local chiralities.

\begin{figure}[tbhp]
\begin{center}
\includegraphics[width=6.5cm,clip]{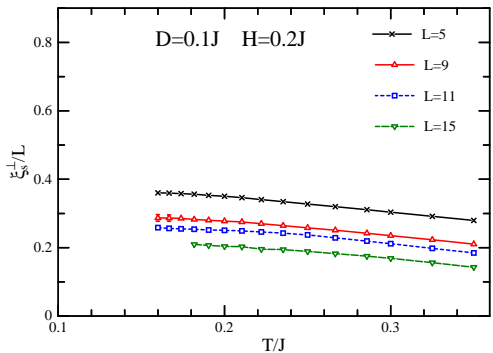}\\
\end{center}
\caption{ (Color online) 
Plots of $\xi_L^\perp /L$ of an anisotropic $\pm J$ model 
with $D = 0.1 J$ at $H=0.2 J$.
}
\label{fig:6}
\end{figure}


In summary, we have examined the phase transition of the three-dimensional 
$\pm J$ Heisenberg models at finite magnetic fields $H \neq 0$.
When anisotropies are absent, results suggest the occurrence of 
the SG (GT) phase transition as in those at $H = 0$. 
On the other hand, no phase transition occurs when they are present. 
These results support the SG scenario of the phase transition of 
Heisenberg SG models. 

Finally, we should note that present results seem also to be compatible 
with experimental observations. 
Petit {\it et al.} performed torque experiments of a series of Heisenberg 
SGs with different magnitudes of local random anisotropies 
in magnetic fields $H$\cite{Petit}. 
They found an irreversibility of the transverse spin components below a 
finite temperature {\it $T_i(H)$}, i.e., a putative SG transition temperature. 
They showed that, when the magnitude of the anisotropy is very weak, 
{\it $T_i(H)$} is almost independent of $H$ at low magnetic fields. 
It is suppressed strongly at $H \neq 0$ as the magnetude is increased. 
If {\it $T_i(H)$} is a crossover temperature between the paramagnetic-like 
state and the SG-like state\cite{Comm_Random}, our present results could 
explain these observations.

\begin{figure}[tbhp]
\begin{center}
\includegraphics[width=5.5cm,clip]{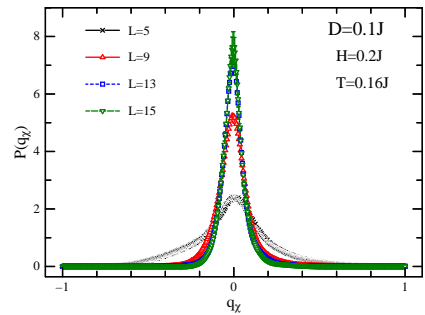}\\
\end{center}
\caption{ (Color online) 
The chiral-overlap distribution functions at $T = 0.16 J$ 
for different lattice sizes $L$. 
}
\label{fig:7}
\end{figure}


\bigskip 

This work was financed by a Grant-in-Aid for Scientific Research 
from Ministry of Education, Culture, Sports, Science and Technology.


\begin{thebibliography}{99}
\bibitem{Comm_MC} In this case, results scatter rather strongly and 
        the equilibration condition is relaxed. Therefore, results 
        for the former $1/2$ MC steps and those for the latter 1/2 MC 
        steps coincide within an error of 5\%. However, results shown 
        in Figs. 6 and 7 are clear-cut.
\bibitem{Comm_q} 
        A scalar chiral-overlap $q_{\chi}$ between the chiralities 
        of the original spins of the two replicas, $\{\bm{S}_i^{\alpha}\}$ 
        and $\{\bm{S}_i^{\beta}\}$, is given in Ref. 17. Here we calculate 
        it using the SG components $\{\tilde{\bm{S}}_i^{\alpha}\}$ 
        and $\{\tilde{\bm{S}}_i^{\beta}\}$.
\end{thebibliography}
\end{document}